\documentclass[3p,times,twocolumn]{elsarticle}

\usepackage{ecrc}

\volume{00}

\firstpage{1}

\journalname{Nuclear Physics B Proceedings Supplement}

\runauth{M.~Dvornikov}

\jid{nuphbp}

\jnltitlelogo{Nuclear Physics B Proceedings Supplement}

\usepackage{amsfonts,amsmath,amssymb,bm,graphicx,subfigure}
\usepackage[figuresright]{rotating}

\begin{document}

\begin{frontmatter}




\title{Generation of cosmic magnetic fields in electroweak plasma}


\author[a,b]{Maxim Dvornikov}

\ead{maxim.dvornikov@usp.br}

\address[a]{Institute of Physics, University of S\~{a}o Paulo, CP 66318, CEP 05315-970 S\~{a}o Paulo, SP, Brazil}

\address[b]{Pushkov Institute of Terrestrial Magnetism, Ionosphere
and Radiowave Propagation (IZMIRAN),
142190 Troitsk, Moscow, Russia}

\begin{abstract}
We study the generation of strong magnetic fields in magnetars and in the early universe. For this purpose we calculate the antisymmetric contribution to the photon polarization tensor in a medium consisting of an electron-positron plasma and a gas of neutrinos and antineutrinos, interacting within the Standard Model. Such a contribution exactly takes into account the temperature and the chemical potential of plasma as well as the photon dispersion law in this background matter. It is shown that a nonvanishing Chern-Simons parameter, which appears if there is a nonzero asymmetry between neutrinos and antineutrinos, leads to the instability of a magnetic field resulting to its growth. We apply our result to the description of the magnetic field amplification in the first second of a supernova explosion. It is suggested that this mechanism can explain strong magnetic fields of magnetars. Then we use our approach to study the cosmological magnetic field evolution. We find a lower bound on the neutrino asymmetries consistent with the well-known Big Bang nucleosynthesis bound in a hot universe plasma. Finally we examine the issue of whether a magnetic field can be amplified in a background matter consisting of self-interacting electrons and positrons.
\end{abstract}


\end{frontmatter}

% \linenumbers


The origin of magnetic fields ($B$ fields) in some astrophysical and cosmological media is still a puzzle for the modern physics and astrophysics. There are multiple models for the generation of strong $B$ fields in magnetars~\cite{DunTho92}. The observable galactic $B$ field can be a remnant of a strong primordial $B$ field existed in the early universe~\cite{Neronov:1900zz}. Recently the indication on the existence of the inflationary $B$ field was claimed basing on the analysis of BICEP2 data~\cite{Bon14}. In the present work we analyze the possibility for the strong $B$ field generation in an electroweak plasma. First, we study the $B$ field generation driven by neutrino asymmetries. Then, we apply our results for the description of strong $B$ fields in magnetars and in the early universe. Finally, we analyze the evolution of a $B$ field in a self-interacting electron-positron plasma.

To study the $B$ field evolution we start with the analysis of the electromagnetic properties of an electroweak plasma consisting of electrons $e^-$, positrons $e^+$, neutrinos $\nu$, and antineutrinos $\bar{\nu}$ of all types. These particles are supposed to interact in frames of the Fermi theory. This interaction is parity violating. Thus the photon polarization tensor $\Pi_{\mu\nu}$ acquires a contribution~\cite{Mohanty:1997mr},
\begin{equation}\label{poltensdec}
  \Pi_{ij}(k) =
  \mathrm{i}\varepsilon_{ijn}k^n \Pi_2+\dots,
\end{equation}
where $\Pi_2 = \Pi_2(k)$ is the new form factor, or the Chern-Simons (CS) parameter, we will be looking for and $k^\mu = (k_0, \mathbf{k})$ is the photon momentum. Here we adopt the notation of~\cite{BoyRucSha12}

\begin{figure}
  \centering
  \includegraphics[scale=.8]{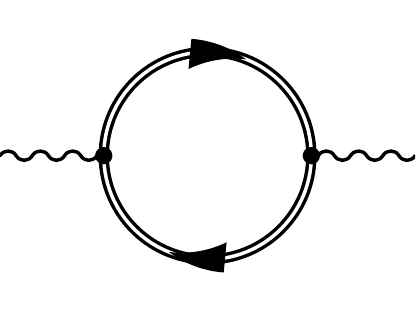}
  \caption{The Feynman diagram for the one loop contribution to the photon
  polarization tensor
  in case of a $e^- e^+$ plasma interacting with a $\nu\bar{\nu}$ gas.
  The electron propagators are shown as broad
  straight lines since they account for the densities of background
  $\nu$ and $\bar{\nu}$~\cite{DvoSem14}.\label{fig:FeynDiagr}}
\end{figure}

First, we will be interested in the contribution to $\Pi_2$ arising from the interaction of a $e^- e^+$ plasma with a $\nu \bar{\nu}$ gas. In this case the most general analytical expression for $\Pi_2$ can be obtained on the basis of the Feynman diagram shown in Fig.~\ref{fig:FeynDiagr}. We shall represent $\Pi_2$ as $\Pi_2 = \Pi_2^{(\nu)} + \Pi_{2}^{(\nu e)}$, where $\Pi_2^{(\nu)}$ is the contribution of only the neutrino gas and $\Pi_{2}^{(\nu e)}$ is the contribution of the $e^- e^+$ plasma with the nonzero temperature $T$ and the chemical potential $\mu$.

The expression for $\Pi_2^{(\nu)}$ can be obtained using the standard quantum field theory technique~\cite{DvoSem14},
\begin{equation}\label{eq:Pivacexpl}
  \Pi_2^{(\nu)} =
  V_5
  \frac{e^{2}}{2\pi^{2}}\frac{k^{2}}{m^{2}}
  \int_{0}^{1}\mathrm{d}x\frac{x(1-x)}{1-\frac{k^{2}}{m^{2}}x(1-x)}.
\end{equation}
where $e$ is the electron charge, $m$ is the electron mass, $V_5 = (V_\mathrm{R} - V_\mathrm{L})/2$, and $V_\mathrm{R,L}$ are the potentials of the interaction of right and left chiral projections of the $e^- e^+$ field with the $\nu \bar{\nu}$ background. The explicit form of $V_\mathrm{R,L}$ can be found in~\cite{DvoSem14}.

The expression for $\Pi_{2}^{(\nu e)}$ can be obtained using the technique for the summation over the Matsubara frequencies~\cite{DvoSem14},
\begin{align}\label{Pi2Tgen}
  \Pi_{2}^{(\nu e)} = &
  V_5 e^{2}
  \int_{0}^{1}dx
  \int\frac{\mathrm{d}^{3}p}{(2\pi)^{3}}
  \frac{1}{\mathcal{E}_{\mathbf{p}}^{3}}
  \notag
  \\
  & \times
  \bigg\{
    I_0^{+} -
    (1-x)
    \bigg[
    \frac{1}{\mathcal{E}_{\mathbf{p}}^2}
    \Big(
      \mathbf{p}^{2}
      \left[
        3 - 5 x
      \right]
      \notag
      \\
      & -
      3x
      \left[
        k^{2}x(1-x)+m^{2}
      \right]
    \Big)
    \left(
      J_{0}^{+}+J_{0}^{-}
    \right)
    \notag
    \\
    & -
    \frac{\beta k_{0}}{2} x(1-2x)
    \left(
      J_{1}^{+}-J_{1}^{-}
    \right) -
    x
    \left(
      J_{2}^{+}+J_{2}^{-}
    \right)
  \bigg]
  \bigg\},
%  \\
%   & 
\end{align}
where
\begin{align}\label{J012}
  I_0^{+} = &
%  \frac{1}{\exp[\beta(\mathcal{E}_{\mathbf{p}}-\mu_{+})]+1} +
  \frac{1}{\exp[\beta(\mathcal{E}_{\mathbf{p}}+\mu_{+})]+1} +
  \frac{\beta\mathcal{E}_{\mathbf{p}}}{2}
%  \left[
%    \frac{1}{\cosh[\beta(\mathcal{E}_{\mathbf{p}}-\mu_{+})]+1} +
    \frac{1}{\cosh[\beta(\mathcal{E}_{\mathbf{p}}+\mu_{+})]+1}
%  \right]
  \notag
  \\
  & +
  (\mu_{+} \to - \mu_{+}),
  \notag
  \\
  J_{0}^{+} = &
  \frac{1}{\exp[\beta(\mathcal{E}_{\mathbf{p}}+\mu_{+})]+1} +
  %\frac{1}{\exp[\beta(\mathcal{E}_{\mathbf{p}}-\mu')]+1} %+
%  \nonumber
%  \\
%  & +
  \frac{\beta\mathcal{E}_{\mathbf{p}}}{2}
  %\left\{
    \frac{1 + \frac{\beta \mathcal{E}_{\mathbf{p}}}{3}
    \tanh
    \left[
      \frac{\beta}{2}(\mathcal{E}_{\mathbf{p}}+\mu_{+})
    \right]
    }
    {\cosh[\beta(\mathcal{E}_{\mathbf{p}}+\mu_{+})]+1}
    \notag
    \\
    & +
    \left(
      \mu_{+} \to - \mu_{+}
    \right),
%    \frac{1+\frac{\beta \mathcal{E}_{\mathbf{p}}}{3}
%    \tanh
%    \left[
%      \frac{\beta}{2}(\mathcal{E}_{\mathbf{p}}-\mu')
%    \right]
%    }
%    {1+\cosh[\beta(\mathcal{E}_{\mathbf{p}}-\mu')]}
  %\right\},
  \notag
  \\
  J_{1}^{+} = & %-\frac{1}{2}
%  \left\{
    \frac{1+\beta\mathcal{E}_{\mathbf{p}}
    \tanh
    \left[
      \frac{\beta}{2}(\mathcal{E}_{\mathbf{p}}+\mu_{+})
    \right]}
    {\cosh[\beta(\mathcal{E}_{\mathbf{p}}+\mu_{+})]+1} -
    \left(
      \mu_{+} \to - \mu_{+}
    \right)
%    \frac{1+\beta\mathcal{E}_{\mathbf{p}}
%    \tanh[\beta(\mathcal{E}_{\mathbf{p}}-\mu^{\pm})/2]}
%    {1+\cosh[\beta(\mathcal{E}_{\mathbf{p}}-\mu^{\pm})]}
%  \right\}
  ,
  \notag
  \\
  J_{2}^{+} = &
  \frac{1}{\exp[\beta(\mathcal{E}_{\mathbf{p}}+\mu_{+})]+1} +
%  \frac{1}{\exp[\beta(\mathcal{E}_{\mathbf{p}}-\mu')]+1} %+
%  \nonumber
%  \\
%  & +
  \frac{\beta\mathcal{E}_{\mathbf{p}}}{2}
  %\left\{
    \frac{1-\beta\mathcal{E}_{\mathbf{p}}
    \tanh
    \left[
      \frac{\beta}{2}(\mathcal{E}_{\mathbf{p}}+\mu_{+})
    \right]
    }
    {\cosh[\beta(\mathcal{E}_{\mathbf{p}}+\mu_{+})]+1}
    \notag
    \\
    & +
    \left(
      \mu_{+} \to - \mu_{+}
    \right).
%    \frac{1-\beta\mathcal{E}_{\mathbf{p}}
%    \tanh
%    \left[
%      \frac{\beta}{2}(\mathcal{E}_{\mathbf{p}}-\mu')
%    \right]
%    }
%    {1+\cosh[\beta(\mathcal{E}_{\mathbf{p}}-\mu')]}
  %\right\} .
\end{align}
Here $\mathcal{E}_{\mathbf{p}} = \sqrt{\mathbf{p}^2 + M^2}$, $\beta = 1/T$, $\mu_{+} = \mu + k_{0}x$ and $M^2 = m^{2}-k^{2}x(1-x)$. To obtain $J_{0,1,2}^{-}$ in Eq.~\eqref{Pi2Tgen} we should replace $\mu_{+} \to \mu_{-} = \mu - k_{0}x$ in $J_{0,1,2}^{+}$ in Eq.~\eqref{J012}. Note that in Eqs.~\eqref{Pi2Tgen} and~\eqref{J012} we assume that $k^2 < 4 m^2$, i.e. no creation of $e^- e^+$ pairs occurs~\cite{Bra92}.

\begin{figure}
  \centering
  \subfigure[]
  {\label{aeenunu}
  \includegraphics[scale=.55]{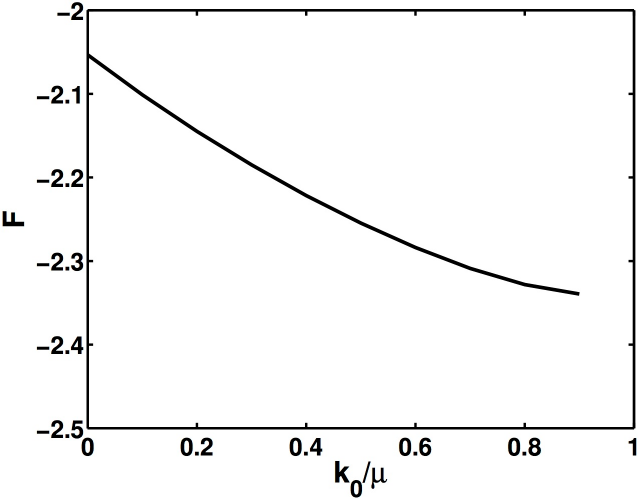}}
  \
  \subfigure[]
  {\label{beenunu}
  \includegraphics[scale=.55]{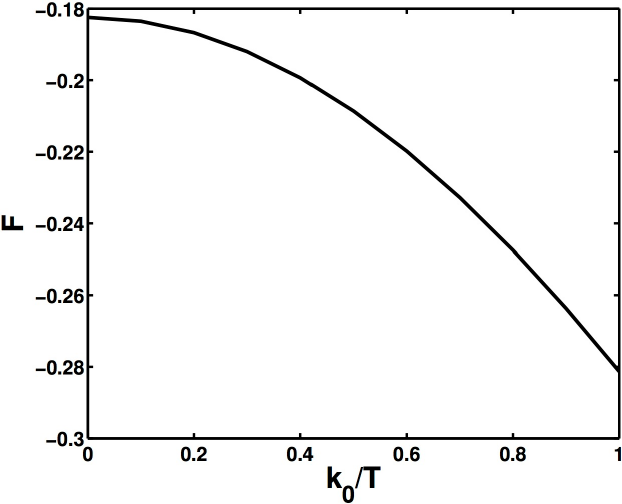}}
  \caption{The function $F$ versus $k_0$ for a $e^- e^+$ plasma interacting
  with a $\nu\bar{\nu}$ gas. (a) Degenerate relativistic plasma.
  (b) Hot relativistic plasma.
  \label{Feenunu}
  }
\end{figure}

It is convenient to represent $\Pi_{2}$ as $\Pi_{2} = 2 \tfrac{\alpha_{\mathrm{em}}}{\pi} V_5 F$, where $F$ is the dimensionless function and $\alpha_\mathrm{em} = \tfrac{e^2}{4\pi}$ is the fine structure constant. Using Eqs.~\eqref{eq:Pivacexpl}-\eqref{J012}, in Fig.~\ref{Feenunu} we show the behavior of $F$ versus $k_0$ in relativistic plasmas. It should be noted that in the static limit $F(k_0 = 0) \neq 0$. To plot Fig.~\ref{Feenunu} we take into account the dispersion law of long electromagnetic waves in plasma $k^2 = k^2(T,\mu)$~\cite{DvoSem14} and the fact that an electron acquires the effective mass $m_\mathrm{eff}^2 = \tfrac{e^2}{8\pi^2} (\mu^2 + \pi^2 T^2)$ in a hot and dense matter~\cite{Bra92}. As shown in~\cite{DvoSem14}, the nonzero $\Pi_2(0) = \Pi_2(k_0=0)$ results in the instability of a $B$ field leading to the exponential growth of a seed field.

We can apply our results for the description of the $B$ field evolution in a dense relativistic electron gas in a supernova explosion. It is known that, just after the core collapse, a supernova is a powerful source of $\nu_e$ whereas the fluxes of $\nu_{\mu,\tau}$ and $\bar{\nu}_{e,\mu,\tau}$ are negligible~\cite{Jan07}. Thus $V_5 \neq 0$ and we get that $\Pi_2(0)=\frac{\sqrt{2}}{\pi}\alpha_\mathrm{em}G_\mathrm{F}n_{\nu_e} F(0) \neq 0$, where $G_\mathrm{F}$ is the Fermi constant and $|F(0)| \approx 2$, see Fig.~\ref{aeenunu}, since electrons are degenerate. The magnetic diffusion time $t_\mathrm{diff} = \sigma \Pi_2^{-2}(0) \approx 2.3 \times 10^{-2}\thinspace\text{s}$ for $n_e = 3.7 \times 10^{37}\thinspace\text{cm}^{-3}$ and $n_{\nu_e} = 1.9 \times 10^{37}\thinspace\text{cm}^{-3}$ in the supernova core~\cite{DvoSem14}. Here $\sigma$ is the electron gas conductivity. Thus at $t \sim 10^{-3}\thinspace\text{s} \ll t_\mathrm{diff}$, when the flux of $\nu_e$ is maximal, no seed magnetic field dissipates. Therefore the neutrino driven instability can result in the growth of the $B$ field. It should be noted that the scale of the $B$ field turns out to be small $\Lambda \sim 10^{-3}\thinspace\text{cm}$. However, at later stages of the star evolution $V_5$ diminishes and $\Lambda$ can be comparable with the magnetar radius. Thus our mechanism can be used to explain strong $B$ fields of magnetars.

Now let us apply out results to study the $B$ field evolution in the primordial plasma. At the stages of the early universe evolution before the neutrino decoupling at $T > (2-3)\thinspace\text{MeV}$, the $e^- e^+$ plasma is hot and relativistic.
Assuming the causal scenario, in which $\Lambda < H^{-1}$, where $H$ is the Hubble constant, we get that $|\xi_{\nu_e} - \xi_{\nu_\mu} - \xi_{\nu_\tau}| > 1.1 \times 10^{-6} \sqrt{g^*/106.75}  \times (T/\mathrm{MeV})^{-1}$, see~\cite{DvoSem14}, where $\xi_{\alpha} = \mu_{\alpha}/T$, $g^*$ is the number of relativistic degrees of freedom, and $\mu_{\alpha}$ is the chemical potential of neutrinos of the type $\alpha = \nu_e,\nu_\mu,\nu_\tau$. Here we use that $|F(0)| \approx 0.2$, see Fig.~\ref{beenunu}. Assuming that before the Big Bang nucleosynthesis at $T\sim (2-3)\thinspace\text{MeV}$ all neutrino flavors equilibrate owing to neutrino oscillations $\xi_{\nu_e} \sim \xi_{\nu_\mu} \sim \xi_{\nu_\tau}$, we get the lower bound on the neutrino asymmetries, which is consistent with the well-known Big Bang nucleosynthesis upper bound on $|\xi_{\alpha}|$, see~\cite{Man12}.

\begin{figure}
  \centering
  \subfigure[]
  {\label{ad}
  \includegraphics[scale=.8]{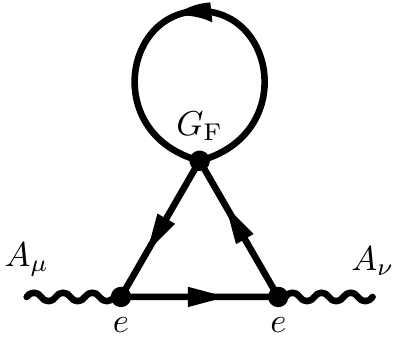}}
  \
  \subfigure[]
  {\label{bd}
  \includegraphics[scale=.8]{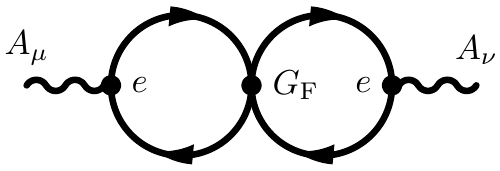}}
  \caption{The Feynman diagrams contributing to the photon
  polarization tensor
  in case of a $e^- e^+$ self-interacting plasma.
  Here $A_\mu$ is the potential of the electromagnetic field.
  \label{2diagr}
  }
\end{figure}

Finally, let us examine the issue of whether a $B$ field can be amplified in a $e^- e^+$ plasma self-interacting within the Fermi model, i.e. when a $\nu\bar{\nu}$ gas is not present. In this case the contributions to $\Pi_2$ are schematically depicted in Fig.~\ref{2diagr}.
The analytical expression for $\Pi_2^{(ee)}$ can be obtained analogously to the previous case~\cite{Dvo14},
\begin{align}\label{eq:P21}
  \Pi_{2}^{(ee)} = &
  \frac{
  \left(
    1-4\sin^{2}\theta_{W}
  \right)
  }
  {2\sqrt{2}}
  e^{2}G_{\mathrm{F}}
  \left(
    n_{e}-n_{\bar{e}}
  \right)
%  \notag
%  \\
%  & \times
  \int_{0}^{1}(1-x)\mathrm{d}x
  \notag
  \\
  & \times
  \int\frac{\mathrm{d}^{3}p}{(2\pi)^{3}}
  \frac{1}{\mathcal{E}_{\mathbf{p}}^{3}}
%  \nonumber
%  \\
%  & \times
  \bigg\{
    \left[
      J'_{2}-J''_{2}
    \right] -
    \left[
      J'_{0}-J''_{0}
    \right]
    \frac{1}{\mathcal{E}_{\mathbf{p}}^{2}}
    \notag
    \\
    & \times
    \left(
      \mathbf{p}^{2}
      \left[
        3 - 2 x
      \right] -
      3
      \left[
        m^{2}(1+x)+k^{2}x^{2}
      \right]
    \right)
  \bigg\},
\end{align}
where $n_{e,\bar{e}}$ are the electron and positron densities, $\theta_{W}$ is the Weinberg angle, and $J'_{0,2} = J^{+}_{0,2}$ in Eq.~\eqref{J012}, with $\mu' = \mu_+$. The expressions for $J''_{0,1}$ can be obtained from $J'_{0,2}$ if we
make the replacement $\mu'\to\mu''=\mu+k_{0}(1-x)$ there. As in deriving of Eqs.~\eqref{Pi2Tgen} and~\eqref{J012}, here we also assume that $k^{2}<4m^{2}$.

\begin{figure}
  \centering
  \subfigure[]
  {\label{aee}
  \includegraphics[scale=.19]{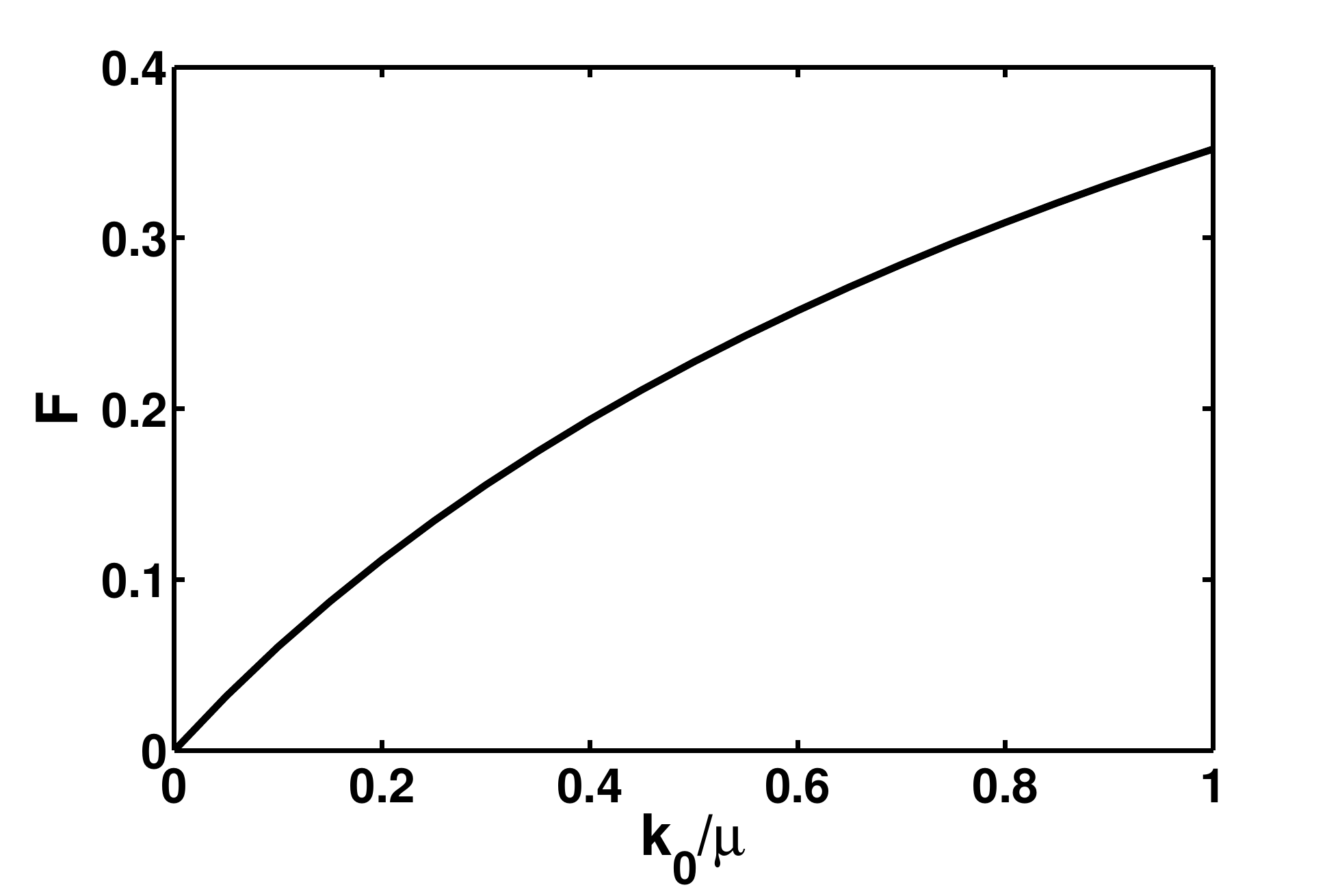}}
  \hskip-.2cm
  \subfigure[]
  {\label{bee}
  \includegraphics[scale=.19]{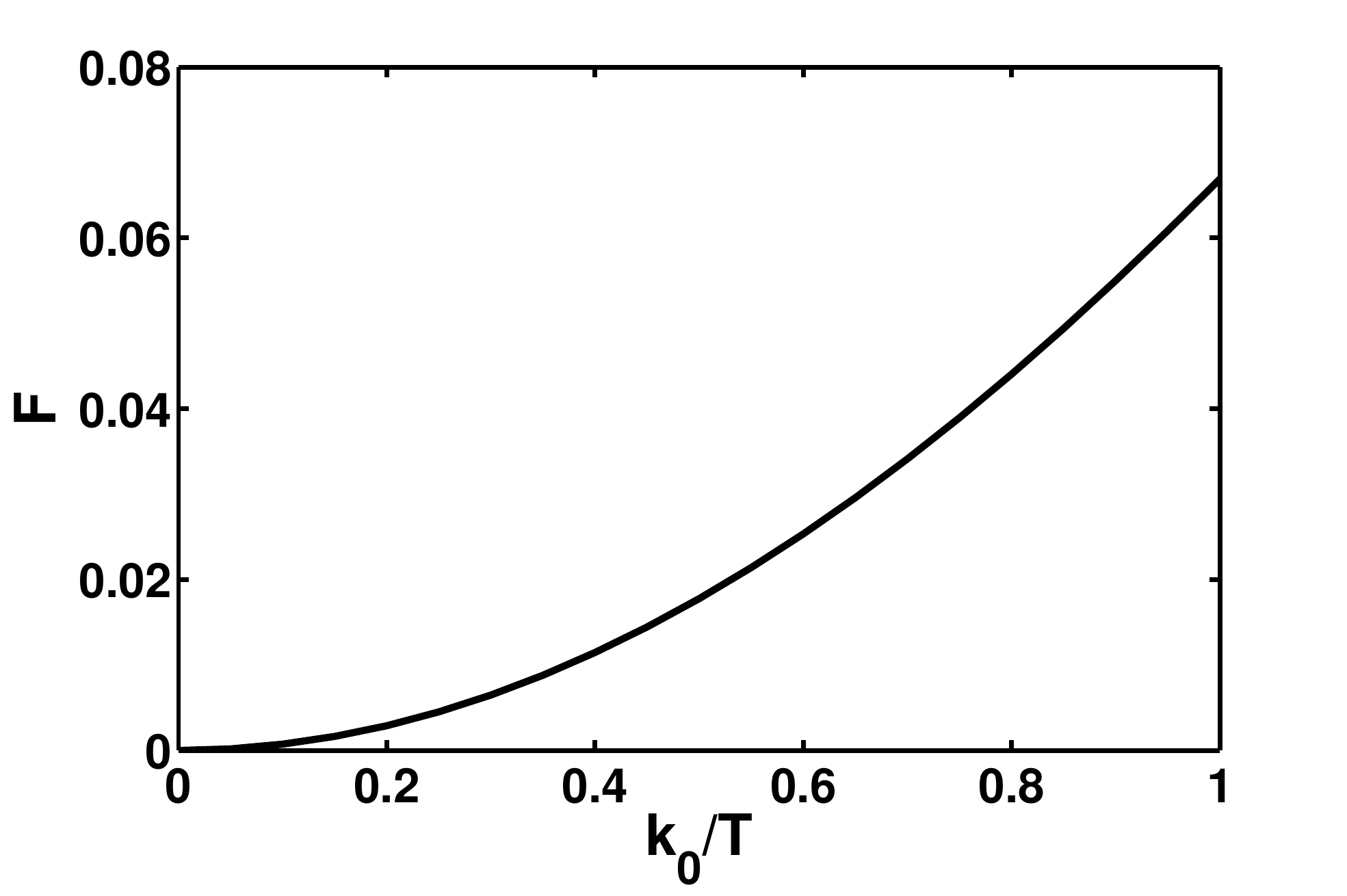}}
  \caption{The function $F$ versus $k_0$ for
  a $e^- e^+$ self-interacting plasma. (a) Degenerate relativistic plasma.
  (b) Hot relativistic plasma.
  \label{Fee}
  }
\end{figure}

Let us express $\Pi_{2}$ in Eq.~(\ref{eq:P21}) as $\Pi_{2}=\tfrac{\alpha_{\mathrm{em}}}{\sqrt{2}\pi} \left( 1-4\sin^{2}\theta_{W} \right) G_{\mathrm{F}} \left( n_{e}-n_{\bar{e}} \right) F$,
where $F$ is the dimensionless function. We shall analyze
this function in the static limit $k_{0}\to0$. We mention that, if
we neglect $k_{0}$ in Eq.~(\ref{eq:P21}), then $J_{0,2}'=J_{0,2}''$
and $\Pi_{2}\to0$. The behavior of $F$ for relativistic plasmas is shown in Fig.~\ref{Fee}, where one can see that $\Pi_{2}(0)=0$. In Fig.~\ref{Fee} we also account for the thermal corrections to the photon dispersion and to the electron mass. It means that a $e^- e^+$ plasma does not reveal the instability of a $B$ field leading to its growth. Therefore, contrary to the claim of~\cite{BoyRucSha12}, one can use this mechanism for neither the explanation of strong $B$ fields of magnetars nor the $B$ field
amplification in the early universe.

In conclusion we mention that we have derived the CS term $\Pi_2$ in an electroweak plasma consisting of $e^-$ and $e^+$ as well as $\nu$ and $\bar{\nu}$ of all flavors. These particles are involved in the parity violating interaction. It makes possible the existence of a nonzero CS term. In case of a $e^- e^+$ plasma interacting with a $\nu\bar{\nu}$ background, the CS term is nonvanishing in the static limit when $k_0 = 0$. Therefore, a $B$ field becomes unstable in this system. We have shown that a seed field can be exponentially amplified. This feature of an electroweak plasma in question can be used to explain strong $B$ fields of magnetars and to study the evolution of a primordial $B$ field. We have also demonstrated that there is no $B$ field instability in a self-interacting $e^- e^+$ plasma.

I am thankful to the organizers of $37^\text{th}$ ICHEP for the invitation and a financial support, to V.B.~Semikoz for helpful discussions, and to FAPESP (Brazil) for a grant.



%% References

%\nocite{*}
%\bibliographystyle{elsarticle-num}
%\bibliography{martin}



\end{document}